\documentclass[prb,superscriptaddress,showpacs,
               reprint]{revtex4-1}

\usepackage[utf8]{inputenc}
\usepackage{amsmath,amsfonts,amssymb}
\usepackage{pifont}
\usepackage{rotating}
\usepackage{graphicx}    
\usepackage{epstopdf}
\usepackage{color}
\usepackage[caption=false]{subfig}
\usepackage{soul}

\usepackage{hyperref}
\definecolor{dark-red}{rgb}{0.9,0.15,0.15}
\definecolor{dark-blue}{rgb}{0.15,0.15,0.4}
\definecolor{medium-blue}{rgb}{0,0,0.5}
\hypersetup{
    colorlinks, linkcolor={black},
    citecolor={dark-blue}, urlcolor={medium-blue}
}

\begin{document}

\title{Structure, magnetic and transport properties of epitaxial thin films of equiatomic CoFeMnGe quaternary Heusler alloy}

\author{Varun K. Kushwaha}
\email{varun.phy@iitb.ac.in}
\affiliation{Low Temperature Laboratory, Department of Physics, Indian Institute of Technology Bombay, Mumbai 400076, India}

\author{Jyoti Rani}
\affiliation{Low Temperature Laboratory, Department of Physics, Indian Institute of Technology Bombay, Mumbai 400076, India}

\author{C. V. Tomy}
\affiliation{Low Temperature Laboratory, Department of Physics, Indian Institute of Technology Bombay, Mumbai 400076, India}

\author{Ashwin Tulapurkar}
\affiliation{Department of Electrical Engineering, Indian Institute of Technology Bombay, Mumbai 400076, India}


\begin{abstract}
Future spintronics requires the realization of thin film of half-metallic ferromagnets having high Curie temperature and 100\% spin polarization at the Fermi level for potential spintronics applications. In this paper, we report the epitaxial thin films growth of half-metallic CoFeMnGe Heusler alloy on MgO (001) substrate using pulsed laser deposition system, along with the study of structural, magnetic and transport properties. The magnetic property measurements of the thin film suggest a soft ferromagnetic state at room temperature with an in-plane magnetic anisotropy and a Curie temperature well above the room temperature. Anisotropic magnetoresistance (AMR) ratio and temperature dependent electrical resistivity measurements of the thin film indicate the compound to be half-metallic in nature and therefore suitable for the fabrications of spintronics devices.  
\end{abstract}


\date{\today}

\maketitle

\section{Introduction}
\begin{figure}[hbtp]
\centering
\includegraphics[width=0.5\linewidth]{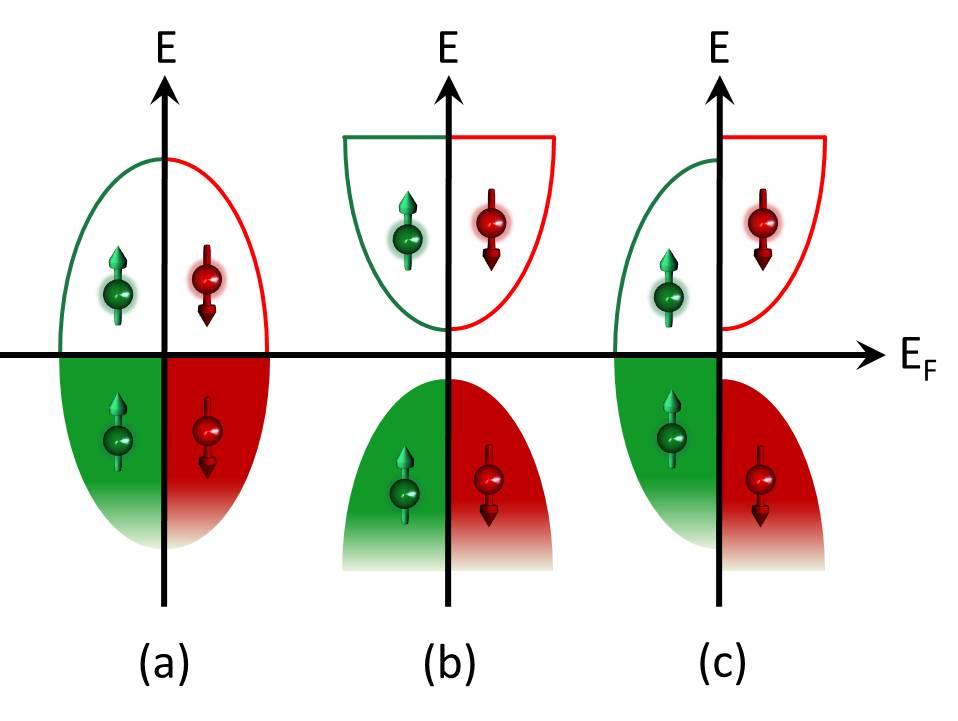}
\caption{Schematic representation of the density of states (DOS) near the Fermi level for a (a) metal, (b) semiconductor and (c) half-metal.}
\label{fig:DOS_CFMG}
\end{figure}

Half-metallic ferromagnets (HMFs), at the Fermi level, possess metallic properties for one spin channel ($\uparrow$) while at the same time semiconducting properties for the other spin channel ($\downarrow$) (see Fig.~\ref{fig:DOS_CFMG}(c)), and thus exhibit 100\% spin polarization. \cite{NiMnSb-deGroot-PRL} Spin injection efficiency in a conventional ferromagnet/semiconductor heterostructure \cite{SpinInjection-PRHammar-PRL} reduces drastically due to the conductivity mismatch \cite{CMismatch-GSchmidt-PRB} and very low spin polarization of the magnetic material. Due to this limiting factor, half-metallic ferromagnet-based HMF/semiconductor heterostructures have been proposed for amplifying the spin injection efficiency. \cite{NiMnSb-deGAWijs-PRB, EfficientSI-SChadov-PRL} As a result, HMFs have attracted much attention in the field of spintronics as one of the most promising candidates for spin injection in spintronics devices. \cite{NiMnSb-deGroot-PRL} Figure~\ref{fig:DOS_CFMG} shows the schematic density of states (DOS) near the Fermi level for a typical metal (a), semiconductor (b), and half-metal (c).

Among the known half, full, and quaternary Heusler alloys, Co-based Heusler alloys \cite{SimplerulesHeusler-TGraf-PSSC,Co2FeSi-SWurmehl-PRB,Co2MnSi-YSakuraba-APL} in particular have attracted much attention for spintronics applications (especially their properties such as giant magnetoresistance (GMR), spin valve, tunnel magnetoresistance (TMR), etc.)\cite{Spintronics-Wolf-Science,Spintronics-SDasSarma-RMP} because of their high Curie temperature (${T_C}$), high saturation magnetization ($M_s$) and 100\% spin polarization. Equiatomic quaternary Heusler alloys (EQHA) XX$^{'}$YZ (where X, X$^{'}$, Y are the transition elements and Z is a $p$-block element, all with 1:1:1:1 stoichiometry) crystallize in Y-type structure (space group F$\overline{4}$3m, \# 216). \cite{SimplerulesHeusler-TGraf-PSSC} Recently, Co-based quaternary Heusler alloys CoFeMn$Z$ ($Z$ = Al, Ga, Si, Ge) \cite{CoFeMnZ-VAlijani-PRB,CoFeMnZ-PKlaer-PRB} were predicted to exhibit half-metallic ferromagnetism with a high $T_C$ ($>$\,550~K) and 100\% spin polarization by ab-initio electronic structure calculations. Specifically, CoFeMnGe (CFMG) can be a potential candidate for efficient spin injection because of its large band gap ($E_g = 2.6$~eV) as compared to other half-metallic Heusler alloys such as Co$_2$MnSi ($E_g$ = 0.41 to 0.81~eV) \cite{Co2MnZ-IShoji-JPSJ}, Co$_2$MnGe ($E_g = 0.54$~eV) \cite{Co2MnX-SPicozzi-PRB} for its minority spin band ($\downarrow$), etc. The half-metallic nature of the bulk CFMG alloy has been determined using point contact Andreev reflection (PCAR) technique at 4~K, \cite{CoFeMnGe-Lakhan-JAP} where the observed spin polarization ($P$) at ${E_F}$ was 70\%, which is larger than those measured in bulk alloys of Co$_2$CrAl ($P$ = 62\%), \cite{Co2CrAl-Karthik-ActaMat} CoFeCrAl (67\%), \cite{CoFeCrAl-Lakhan-JMMM} $L2_1$-ordered Co$_2$Mn(Si, Ge) (54\%, 58\%) thin films, \cite{Co2MnGe-Co2MnSi-Rajanikanth-JAP} and comparable with CoFeCrAl (68\%), \cite{CoFeCrAl-YJin-APL} and Co$_2$FeGa$_{0.5}$Ge$_{0.5}$ (75\%) \cite{Co2FeGaGe-Varaprasad-ActaMat} thin films. In 2012, Kokado \textit{et al.} \cite{AMR-Kokado-JPSJ} reported that the sign of anisotropic magnetoresistance (AMR) effect can also be used as a signature of half-metallicity/non-half-metallicity in magnetic materials.

The change in the electrical resistivity as a function of relative angle between the  current density ($\overrightarrow{{J}}$)  and the magnetization ($\overrightarrow{{M}}$) directions is called as the  anisotropic magnetoresistance (AMR). \cite{AMR-Kokado-JPSJ, AMR-Malozemoff-PRB} The AMR ratio is generally defined as:
\begin{equation}
\frac{\Delta\rho}{\rho} =  \frac{\rho_{\parallel} - \rho_{\perp}}{\rho_{\perp}}
\end{equation}             
where $\rho_{\parallel}$  and $\rho_{\perp}$ are the resistivities for $\overrightarrow{{M}}\Vert\overrightarrow{{J}}$ and $\overrightarrow{{M}}\bot\overrightarrow{{J}}$, respectively. According to the theoretical prediction of Kokado \textit{et al.} \cite{AMR-Kokado-JPSJ}, a negative AMR ratio is the signature of  half-metallicity in magnetic materials. This is because when the dominant $s$-$d$ scattering process occurs between the $s$- and the $d$- states which have the same type of spins (s$\uparrow$ $\rightarrow$ d$\uparrow$ or s$\downarrow$ $\rightarrow$ d$\downarrow$ ), the AMR ratio is negative ($\rho_{\parallel} < \rho_{\perp}$), and when the dominant scattering occurs between the opposite spin states (s$\uparrow$ $\rightarrow$ d$\downarrow$ or s$\downarrow$ $\rightarrow$ d$\uparrow$), the sign tends to be positive.

In this paper, we report the fabrication of epitaxial thin films of the CFMG alloy, which in the bulk form has a Curie temperature ${T_C}$ and saturation magnetization ${M_s}$ of 711~K and 3.8~$\mu_{B}\mathrm{/f.u.}$ (which is close to the Slater-Pauling value of 4~$\mu_{B}\mathrm{/f.u.}$), respectively, \cite{CoFeMnZ-VAlijani-PRB,CoFeMnZ-PKlaer-PRB} followed by the study of their structural, magnetic and transport properties. The structural analyses confirm the epitaxial nature of the films, the magnetic measurements indicate soft ferromagnetism with in-plane anisotropy and the transport  measurements show the fingerprint of half-metallicity.

\section{Experimental details}
The thin films of CFMG in this study were prepared using pulsed laser deposition (PLD) system on a MgO (001) single crystal substrate. MgO substrate was used for reducing the lattice mismatch ($a_{\mathrm{CFMG}}-\sqrt{2}a_{\mathrm{MgO}})/\sqrt{2}a_{\mathrm{MgO}} \approx -3.2$\%) between the MgO and the CFMG film. Before the thin film deposition, the MgO substrate was annealed at 750${^\circ}$C for 30 minutes to remove the surface contamination, if any. A KrF excimer laser (wavelength $\lambda = 248$~nm) was used to ablate the target of the CFMG alloy. The laser repetition rate was 5~Hz, the target to substrate distance was fixed at 4~cm, base pressure was $1.8\times10^{-6}$~mbar and the areal energy density was fixed at 3~J/cm$^{2}$. We deposited the CFMG thin films at various deposition temperatures (${T_D}$) and each deposited thin film was annealed \textit{in-situ} at 700${^\circ}$C for 1 h to enhance the crystallization and chemical ordering. 

The structure of the CFMG thin films and their thickness were analysed by the X-ray diffraction (XRD) and the X-ray reflectivity (XRR) measurements, respectively, using a high resolution X-ray diffractometer with Cu-K$_\alpha$ radiation ($\lambda = 1.5406$~\AA). The surface topography and the atomic ratio of the films were determined using the atomic force microscope (AFM) and the energy dispersive X-ray spectroscopy (EDS), respectively. The temperature ($T$) and the magnetic field ($H$) dependence of magnetization ($M$) was carried out using a superconducting quantum interference device-vibrating sample magnetometer (SQUID-VSM, Quantum Design, USA). A physical property measurement system (PPMS, Quantum Design, USA) was used for measuring the electrical resistivity ($\rho$) and anisotropic magnetoresistance (AMR) using four-probe method by applying an ac current.

\begin{figure}[hbtp]
\centering
\includegraphics[width=0.9\linewidth]{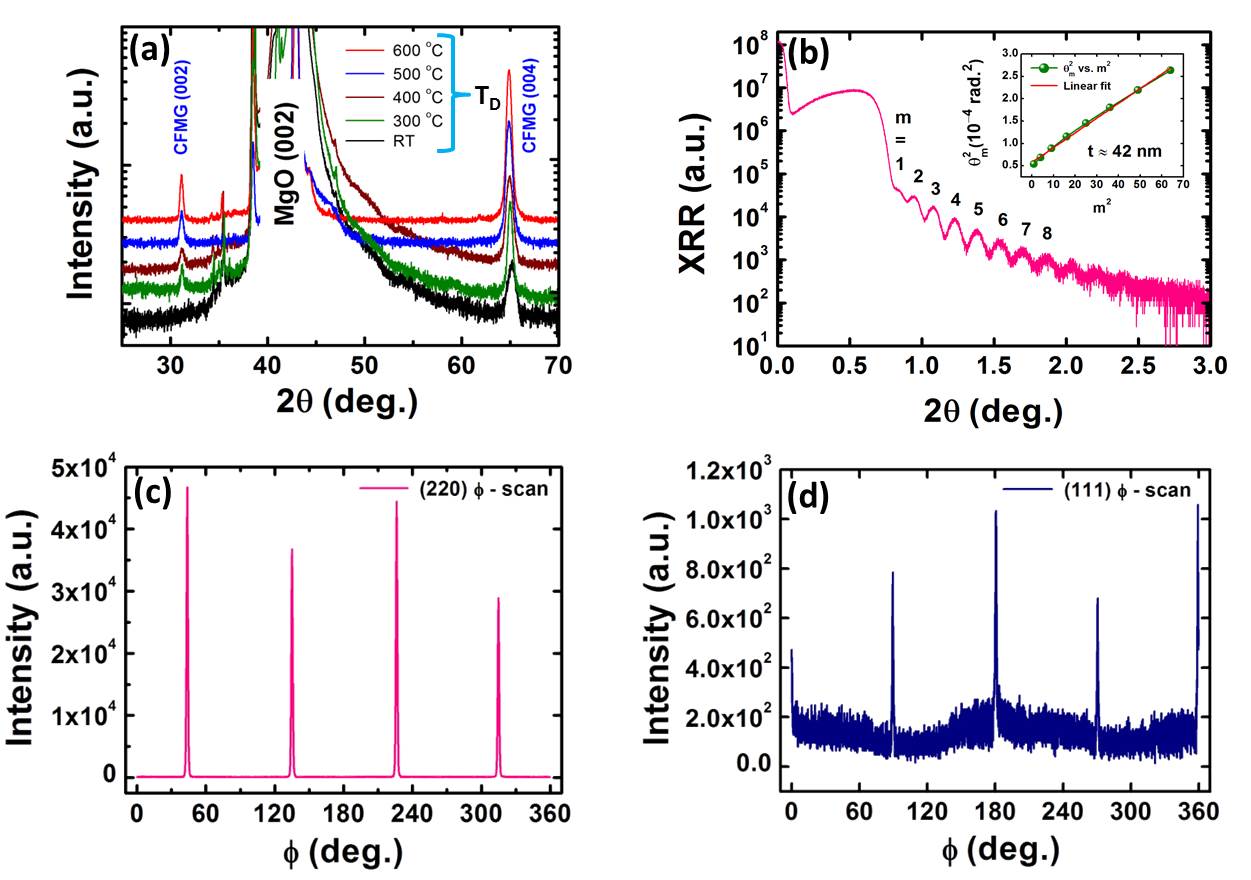}
\caption{(a) XRD $\omega-2\theta$ (out-of-plane) of CFMG thin films deposited on MgO substrates. (b) XRR data and its linear fit for calculating the film thickness. Out-of-plane $\phi$-scan of: (c) (220) and (d) (111) peaks.}
\label{fig:XRD_CFMG}
\end{figure}

\section{Results and discussions}
\subsection{Structural analysis}
\begin{figure}[hbtp]
\centering
\includegraphics[width=0.5\linewidth]{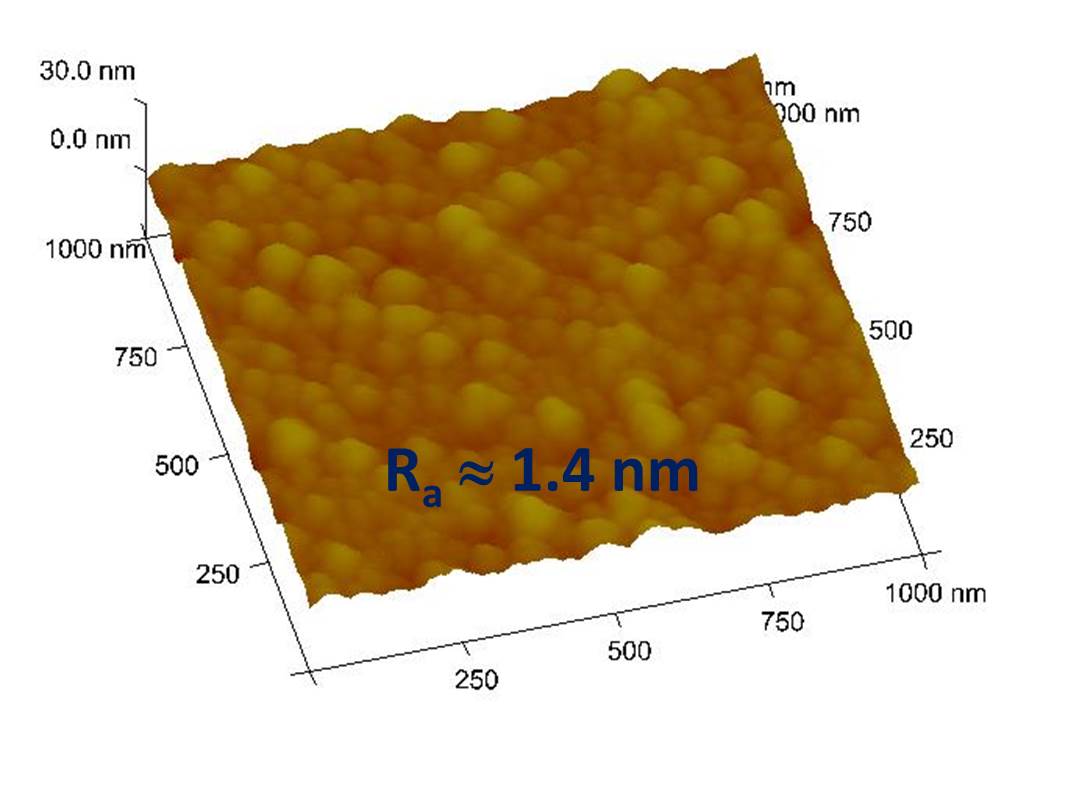}
\caption{AFM image of thin film surface over 1\,$\mu$m~$\times$~1\,$\mu$m area at room temperature.}
\label{fig:AFM_CFMG}
\end{figure}

Figure~\ref{fig:XRD_CFMG}(a) shows the room temperature  X-ray diffraction pattern of the CFMG thin films deposited on MgO (001) substrate at various deposition temperatures (${T_D}$ = RT (room temperature) to 600${^\circ}$C). Only the (002) and (004)  diffraction peaks are present in our CFMG film other than the peaks from the MgO substrate, indicating the epitaxial growth along the (001) direction.  We have also observed that the increase in deposition temperature enhances the intensity of the (002) and (004) diffraction peaks, with the optimum  crystallinity for the film grown at  ${T_D} = 600{^\circ}$C. Therefore we have used only the thin film deposited at ${T_D} = 600{^\circ}$C for further studies; structural, magnetic and transport properties analysis. The lattice constant $a$ of the CFMG film was estimated to be around 5.74~\AA, which compares well with the reported experimental value (5.76~\AA) \cite{CoFeMnZ-VAlijani-PRB} for the bulk CFMG sample.  In addition, the atomic ratio of the the film determined by EDS was close to 1:1:1:1 (with $\pm$2\% for each element), indicating the formation of the stoichiometric compound. The thickness of the thin film was calculated from the XRR spectra, which is shown in Fig.~\ref{fig:XRD_CFMG}(b) and its inset, using the modified Bragg equation: \cite{Book-Birkholz-X-ray, CoFeMnSi-Varun-APL}

\begin{equation}
\theta_m^2 = \theta_c^2 + \left(\frac{\lambda}{2t}\right)^2m^2
\end{equation}
\noindent where $m$ is an integer (fringe order), $\lambda$ is the X-ray wavelength of Cu-K$_\alpha$ radiation, $t$ is the film thickness, $\theta_c$ is the critical angle (in radians) of incidence and $\theta_m$ is the Bragg angle (in radians) of the $m$\textsuperscript{th} oscillation maxima. The calculated thickness was found to be around 42~nm. Figure~\ref{fig:AFM_CFMG} shows the surface topography  obtained from the AFM scans for an area of 1\,$\mu$m~$\times$~1\,$\mu$m  of the thin film at room temperature. The measured average roughness (${R_a}$) for this film is around 1.4~nm.

The epitaxial nature of the film was further confirmed by measuring the four-fold symmetry of (220) peaks (2$\theta = 44.59{^\circ}$, $\chi = 45{^\circ}$) in the $\phi$-scan measurement (Fig.~\ref{fig:XRD_CFMG}(c)). The $\phi$-scan of the (111) peak (2$\theta$ = 26.87${^\circ}$, $\chi$ = 54.7${^\circ}$)  also displays a four-fold symmetry (Fig.~\ref{fig:XRD_CFMG}(d)), indicating a ${Y}$ or an ${L2_1}$-type  structure\cite{SimplerulesHeusler-TGraf-PSSC} for the CFMG film. Furthermore, the diffraction peaks (220) and (111) in the $\phi$-scan are separated by an angle of 45${^\circ}$, which is expected for the cubic crystal structure of the CFMG alloy. Since the scattering factors ($f$) of Co  and Fe  are almost similar, it is practically impossible to distinguish between the  $L2_1$- and the $Y$-type structures using the X-ray Cu-K$_\alpha$ radiation and hence one needs special measurements like neutron-diffraction. We have also made an attempt to estimate the values of the long-range-parameters, $S_{B2}$ and $S_{L2_1}$ using the extended Webster model\cite{L21-order-Takamura-JAP} since these parameters can provide some idea about the  degree of $L2_1$ and $B2$ ordering in the structure.  The values obtained, $S_{B2} = 2.6$ and $S_{L2_1}=5.1$, are found to be too high compared to the values that is expected (maximum value of 1) for these parameters, indicating that this analysis is not useful to determine the type of structural ordering in the present case.

\subsection{Magnetic properties}
\begin{figure}[hbtp]
\centering
\includegraphics[width=0.7\linewidth]{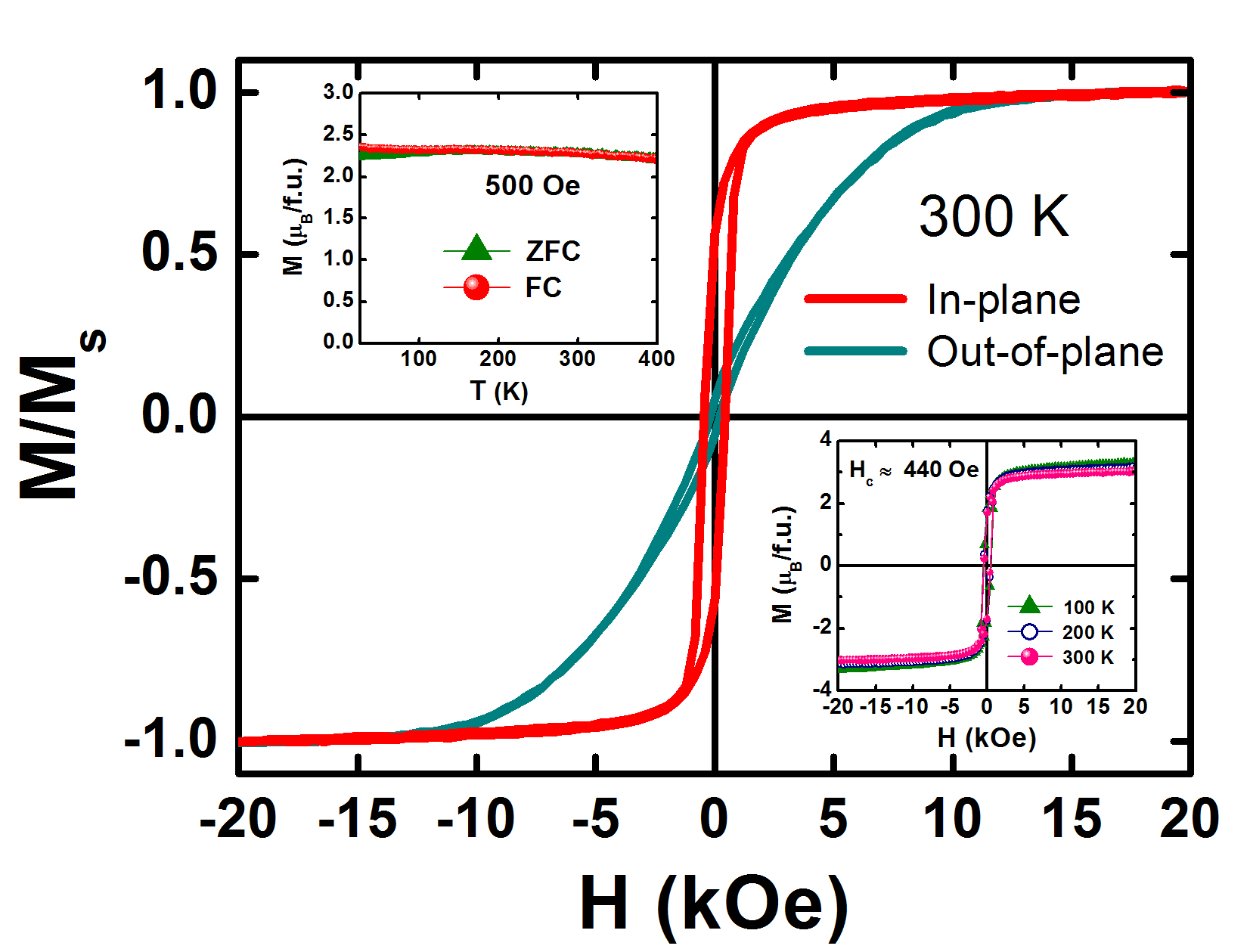}
\caption{The in-plane (IP) and out-of-plane (OP) isothermal magnetization ($M$-$H$) curves of the film as function of applied magnetic field ($H$) at 300~K. The upper inset shows the temperature dependence ($T$) of magnetization ($M$-$T$) curve and the lower inset shows the magnetic field dependence of magnetization ($M$-$H$) curve.}
\label{fig:MTMH_CFMG}
\end{figure}

Figure~\ref{fig:MTMH_CFMG} shows the magnetic properties of the thin film as function of temperature ($T$) and magnetic field ($H$). The temperature dependent magnetization (upper inset) indicates that the ${T_C}$ of the film is well above 400~K. From the magnetization curves (for both in-plane and out-of-pane) as a function of magnetic field (see main panel), it is obvious that the film is ferromagnetically soft along the in-plane (IP) direction with a coercive field of 440~Oe.  The saturation magnetization value is 3.04~$\mu_{B}\mathrm{/f.u.}$ at 300~K (lower inset), which is less than the reported experimental bulk value (3.8~$\mu_{B}\mathrm{/f.u.}$). \cite{CoFeMnZ-VAlijani-PRB, CoFeMnZ-PKlaer-PRB} This discrepancy in magnetic moment may be attributed to the assumption of the same thickness over the entire area of the film (if the thickness is assumed to be 35 nm, the the estimated moment will be 3.64~$\mu_{B}\mathrm{/f.u.}$). Such discrepancies in magnetic moments of thin film samples were also observed in other compounds, e.g. CoFeMnSi. \cite{CoFeMnSi-Varun-APL}

\subsection{Transport properties} 
\begin{figure}[hbtp]
\centering
\includegraphics[width=0.7\linewidth]{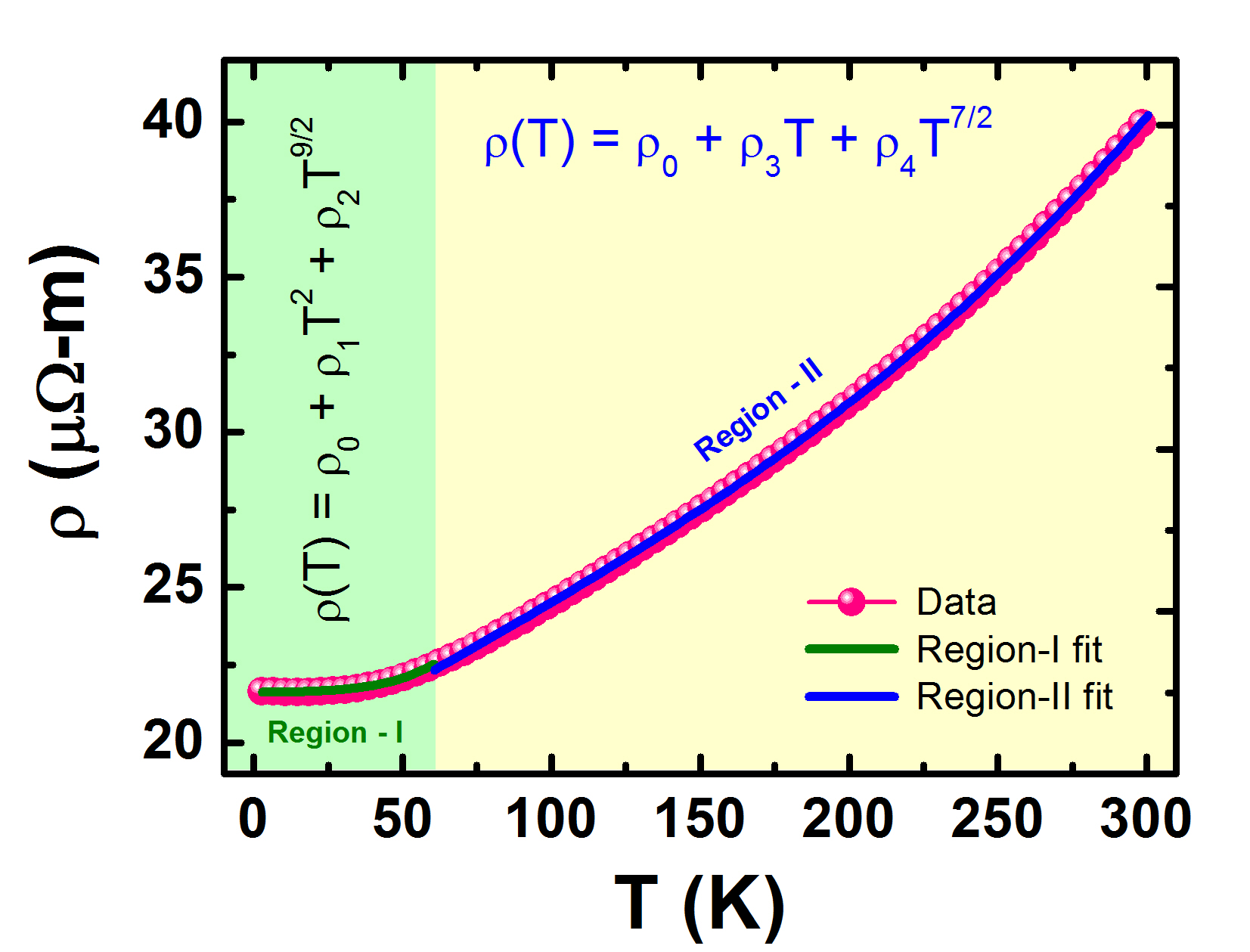}
\caption{Temperature dependence of electrical resistivity in the temperature range of 2-300 K with fitting (solid lines) using two  different equations (one for low and other for high temperature).}
\label{fig:Res_CFMG}
\end{figure}

Figure~\ref{fig:Res_CFMG} shows the variation of electrical resistivity ($\rho$) as function of temperature ($T$). Resistivity of the film increases with increasing temperature from 2-300~K and shows a typical metallic behaviour with a residual resistivity ($\rho_0$) value of $\sim21.62$~$\mu\Omega$-m. The value of the residual resistivity ratio (RRR = $\rho_{300K}$/$\rho_{2K}$ = 1.85) of the film is higher than that of the bulk CFMG (RRR = 1.02) \cite{CoFeMnGe-Lakhan-JAP} and the thin films of some other Heusler alloys such as $\mathrm{Co_2MnSi}$ (RRR = 1.25), \cite{Co2MnSi-LJSingh-APL} $\mathrm{Co_2MnGe}$ (RRR = 1.3), \cite{Co2MnZ-UGeiersbach-JMMM} and $\mathrm{Co_2FeSi}$ (RRR = 1.5). \cite{Co2FeSi-HSchneider-PRB} Resistivity can also be used as an indirect way to estimate the nature of half-metallic character in magnetic materials. Here, we have analysed the temperature dependent resistivity ($\rho(T)$) behaviour of the thin film by considering the possible scattering mechanisms in a half metallic ferromagnet such as (i) electron-phonon scattering (linear $T$-dependence), \cite{Co2MnSi-LJSingh-APL, NiMnSb-SGardelis-JAP} (ii) electron-electron or electron-magnon scattering ($T^2$-dependence), \cite{Co2MnSi-LJSingh-APL, NiMnSb-SGardelis-JAP} and (iii) magnon-magnon scattering ($T^{\frac{9} {2}}$-dependence for low temperature and $T^{\frac{7} {2}}$-dependence  for high temperature). \cite{Book-HMAMM-Kronm-Parkin} In a perfect half-metal, single magnon scattering is not possible due to the energy  gap at the Fermi level for one spin channel. Thus the magnon-magnon scattering is expected to be the dominant scattering mechanism in half metals due to the absence of spin-flip scattering (because of the band gap in the minority spin channel).

We have to divide the resistivity into two regions to obtain the best fit;  the resistivity data in the low temperature region (2-60~K) was fitted using the relation $\rho (T) = \rho_{0} + \rho_{1}T^{2} + \rho_{2}T^{\frac{9}{2}}$, while the remaining data  (60-300~K) was fitted with the relation $\rho (T) = \rho_{0} + \rho_{3} T + \rho_{4}T^{\frac {7}{2}}$.  $\rho_{0}$ is the residual resistivity while  $\rho_{1}$, $\rho_{2}$, $\rho_{3}$, and $\rho_{4}$ are the fitting parameters corresponding to different scattering mechanisms. Values of the fitting parameters are listed in table I. It is obvious from Fig.~\ref{fig:Res_CFMG} that the low temperature region (Region I) of the resistivity data fits well to the $T^{\frac{9}{2}}$ dependence while a $T^{\frac {7}{2}}$ dependence is more suited in the high temperature region (Region II) of the  $\rho (T)$ curve. Such a dependence of the resistivity clearly suggests the half-metallic nature of the compound in the thin film form. 

\begin{table}[h]
\caption {parameters obtained from the best fit of the resistivity data} 
\begin{tabular}{|c|c||c|c|} 
\hline
\multicolumn{2}{|c||}{Region I} &\multicolumn{2}{|c|}{Region II} \\ 
\hline
$\rho_{0}$ &21.62 &$\rho_{0}$ &19.12 \\
&$\mu\Omega$-m & &$\mu\Omega$-m \\
\hline
$\rho_{1}$ &7.84$\times10^{-5}$ &$\rho_{3}$ &0.05 \\
&$\mu\Omega$-m/K$^{2}$ & &$\mu\Omega$-m/K \\
\hline
$\rho_{2}$ &6.00$\times10^{-9}$ &$\rho_{4}$ &1.11$\times10^{-8}$ \\
&$\mu\Omega$-m/K$^{\frac{9}{2}}$ & &$\mu\Omega$-m/K$^{\frac{7}{2}}$\\
\hline
\end{tabular}
\end{table}  

\begin{figure}[hbtp]
\centering
\includegraphics[width=0.8\linewidth]{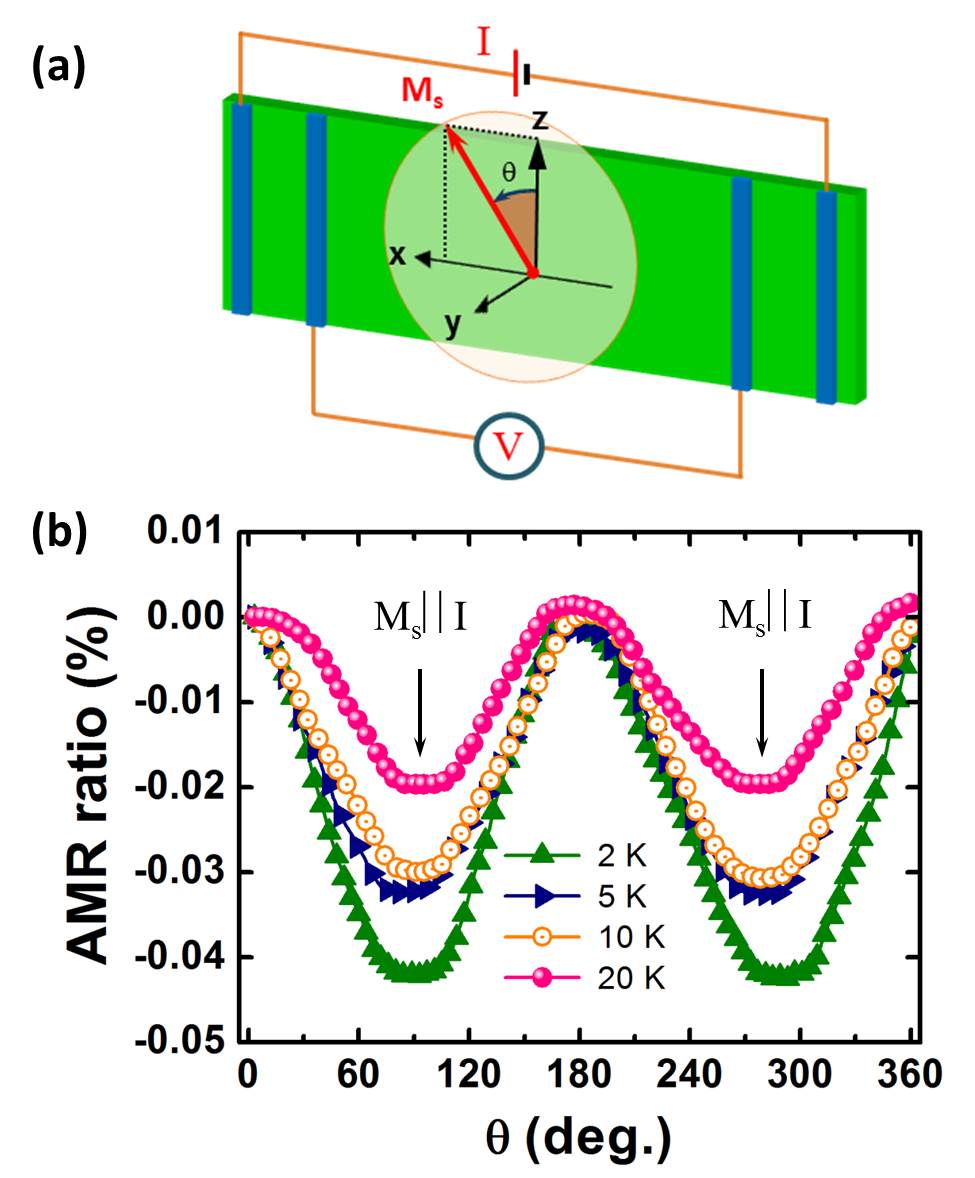}
\caption{(a) The schematic illustration of in-plane AMR measurement, (b) the dependence of AMR ratio on the in-plane relative angle $\theta$ in CFMG thin film at low temperatures.}
\label{fig:AMR_CFMG}
\end{figure}

In order to further corroborate  the nature of half-metallicity of the film, in-plane anisotropic magnetoresistance (AMR) of the film was measured. Figure~\ref{fig:AMR_CFMG}(a) shows the schematic illustration of the in-plane AMR measurement where the directions of the current ($I$), the magnetic field ($H$) i.e., saturation magnetization ($M_s$), and the relative angle $\theta$ between $M_s$ and $I$ are clearly indicated. A magnetic field of 80~kOe, which is high enough to obtain the saturation magnetization, was applied along the $z$-axis and the film was rotated in the $x$-$z$ plane after applying an ac current along the $x$-direction.  The dependence of the AMR ratio on the in-plane relative angle $\theta$ in our CFMG thin film at different temperatures is shown in Fig.~\ref{fig:AMR_CFMG}(b). The AMR and thus the AMR ratio show a clear two-fold symmetry, which is expected in magnetic materials. In the absence of PCAR measurement, the negative sign of the AMR ratio provides an additional confirmation that the thin film of CFMG prepared by us is half-metallic which may be due to the s$\uparrow$ $\rightarrow$ d$\uparrow$ dominant $s$-$d$ scattering. The observed AMR ratio of $-$0.02~\% to $-$0.04~\% is almost the same as that reported in the thin film of half-metallic NiMnSb Heusler compound \cite{NiMnSb-AMR-ZWen-SReport} and one order of magnitude smaller than that reported for other half-metallic Heusler thin films. \cite{Co2-FeMn-Si-AMR-Yang-PRB, Co2MnZ-Co2FeZ-AMR-Sakuraba-APL} We can also clearly see that the magnitude of the AMR ratio decreases as the temperature is increased from 2~K to 20~K, which is an indication of  weak half-metallicity against the thermal fluctuations in CFMG thin film.

\section{Conclusions}
In summary, epitaxial (001)-oriented thin films of CFMG quaternary Heusler alloy were grown on single crystal MgO (001) substrates using pulsed laser deposition system. The magnetization measurements confirm that the film is ferromagnetically soft along the in-plane direction and its Curie temperature is well above 400~K. In the absence of point-contact Andreev reflection (PCAR) technique, we have used an alternate method (in-plane AMR ratio measurement) to determine the half-metallic nature of the thin film. Analysis of the temperature dependence of electrical resistivity indicated that the film has half-metallic behaviour, which was also confirmed by the  negative sign of the AMR ratio. To get a clear picture about the spin band gap (i.e., half-metallicity/spin polarization) in these thin films, PCAR or spin-resolved  photoemission spectroscopy (SRPES) measurements will be highly desirable. The signature of half-metallicity and the high Curie temperature ($T_C$) make these films as promising material for efficient spin current injection in spintronics devices (e.g., spin valve, GMR, TMR, etc.). However, further experiments are necessary to know the type of structural ordering and also to achieve robust half-metallicity against the thermal fluctuation in CFMG thin films.


\section*{Acknowledgment}
All the authors are thankful to the Centre of Excellence in Nanoelectronics (CEN) and Nanofabrication facility (IITBNF) at  IIT-Bombay for the thin films preparation and the Institute central facilities (High resolution-XRD, AFM, EDS, and Magnetometer) for characterizations.


\bibliographystyle{apsrev4-1}
\bibliography{bib}
\end{document}